%% file: awsdm2020.tex
\documentclass[sigconf]{acmart}

\AtBeginDocument{%
  \providecommand\BibTeX{{%
    \normalfont B\kern-0.5em{\scshape i\kern-0.25em b}\kern-0.8em\TeX}}}
    
 \setcopyright{none}
\renewcommand\footnotetextcopyrightpermission[1]{} 

\setcopyright{acmlicensed}
\copyrightyear{2020}
\acmYear{2020}
\acmDOI{10.1145/1122445.1122456}

\acmConference[WSDM '20]{WSDM '20: ACM International Conference on Web Search and Data Mining}{February 03--07, 2020}{Houston, TX}
\acmBooktitle{WSDM '20: ACM International Conference on Web Search and Data Mining,
  February 03--07, 2020, Houston, TX}
\acmPrice{15.00}
\acmISBN{978-1-4503-9999-9/18/06}

\usepackage{amsmath,amssymb,amsfonts}
\usepackage{algorithmic}
\usepackage{graphicx}
\usepackage{textcomp}
\usepackage{xcolor}

\usepackage{multirow}
\usepackage{subcaption}
\usepackage{url}
\usepackage{xspace}

\begin{document}

\title{GREASE: A Generative Model for Relevance Search over Knowledge Graphs}

\author{Tianshuo Zhou}
\affiliation{%
  \institution{National Key Laboratory for Novel Software Technology, Nanjing University, China}
}
\email{tianshuo.zhou@smail.nju.edu.cn}

\author{Ziyang Li}
\affiliation{%
  \institution{National Key Laboratory for Novel Software Technology, Nanjing University, China}
}
\email{zyli@smail.nju.edu.cn}

\author{Gong Cheng}
\affiliation{%
  \institution{National Key Laboratory for Novel Software Technology, Nanjing University, China}
}
\email{gcheng@nju.edu.cn}

\author{Jun Wang}
\affiliation{%
  \institution{Department of Computer Science, University College London, UK}
}
\email{j.wang@cs.ucl.ac.uk}

\author{Yu'Ang Wei}
\affiliation{%
  \institution{National Key Laboratory for Novel Software Technology, Nanjing University, China}
}
\email{weiyuang@smail.nju.edu.cn}

\renewcommand{\shortauthors}{Zhou, et al.}

\renewcommand{\algorithmicrequire}{\textbf{Input:}}
\renewcommand{\algorithmicensure}{\textbf{Output:}}

\newcommand{\powerset}{\ensuremath{\mathtt{P}}\xspace}
\newcommand{\rel}{\ensuremath{\mathtt{rel}}\xspace}
\newcommand{\regularization}{\ensuremath{\mathtt{J}}\xspace}
\newcommand{\ans}{\ensuremath{\mathtt{Ans}}\xspace}
\newcommand{\length}{\ensuremath{\mathtt{len}}\xspace}
\newcommand{\prb}{\ensuremath{\mathtt{Pr}}\xspace}
\newcommand{\pc}{\ensuremath{\mathtt{pc}}\xspace}
\newcommand{\apc}{\ensuremath{\mathtt{apc}}\xspace}
\newcommand{\sametype}{\ensuremath{\mathtt{ST}}\xspace}

\newcommand{\gong}[1]{\textcolor{green}{#1}}
\newcommand{\zhou}[1]{\textcolor{purple}{#1}}

\begin{abstract}
Relevance search is to find top-ranked entities in a knowledge graph~(KG) that are relevant to a query entity. Relevance is ambiguous, particularly over a schema-rich KG like DBpedia which supports a wide range of different semantics of relevance based on numerous types of relations and attributes. As users may lack the expertise to formalize the desired semantics, supervised methods have emerged to learn the hidden user-defined relevance from user-provided examples. Along this line, in this paper we propose a novel generative model over KGs for relevance search, named GREASE. The model applies to meta-path based relevance where a meta-path characterizes a particular type of semantics of relating the query entity to answer entities. It is also extended to support properties that constrain answer entities. Extensive experiments on two large-scale KGs demonstrate that GREASE has advanced the state of the art in effectiveness, expressiveness, and efficiency.
\end{abstract}


\maketitle

\input{introduction.tex}
\input{problem.tex}
\input{approach.tex}

\input{experiment.tex}
\input{related_work_icdm.tex}
\input{conclusion.tex}

\bibliographystyle{ACM-Reference-Format}
\bibliography{awsdm2020}

\end{document}

%% file: introduction.tex
\section{Introduction}

\begin{figure}[t]
\centering
\includegraphics[width=\linewidth]{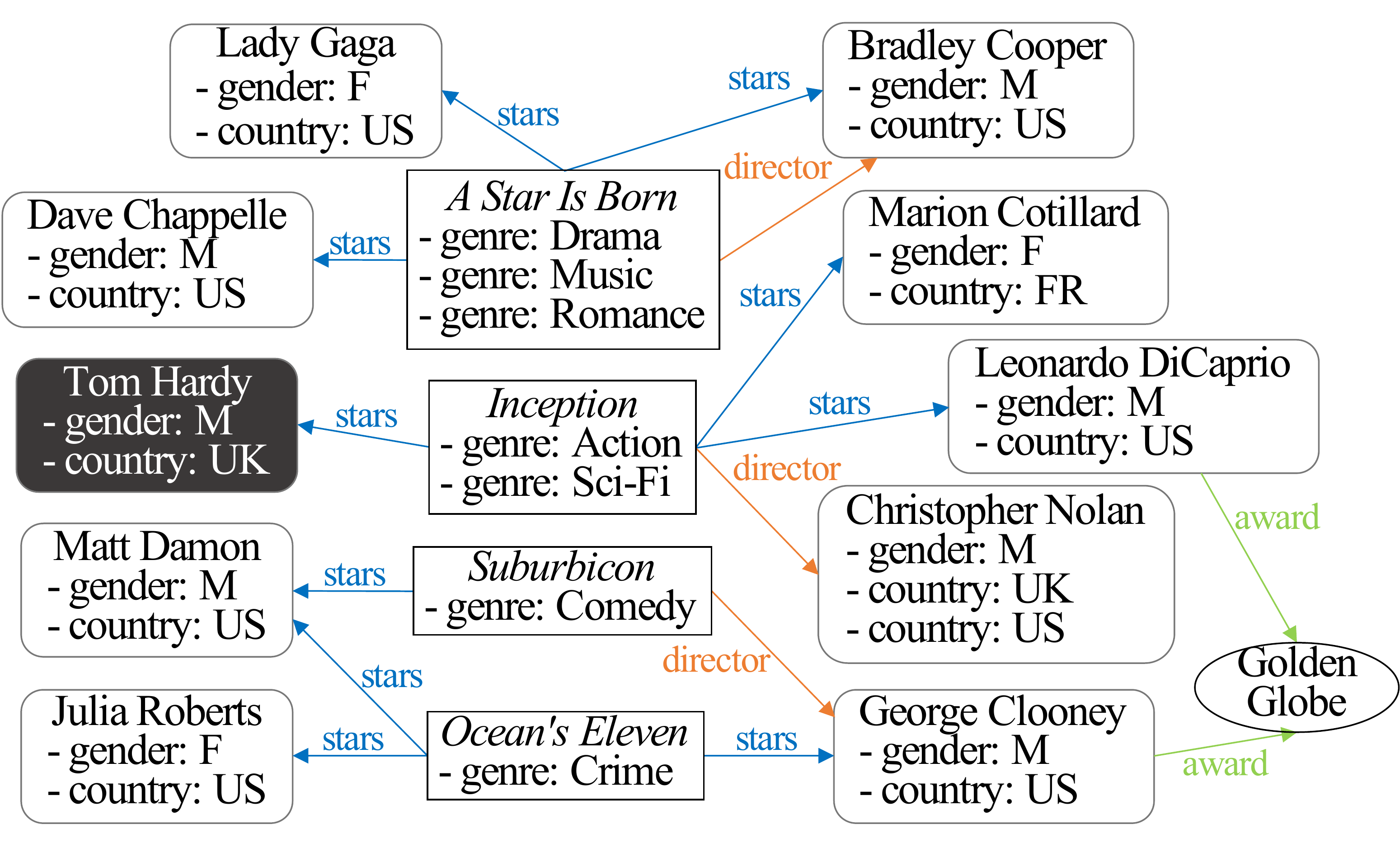}
\caption{An example of a knowledge graph, where each entity is associated with a bulleted set of attributes.}
\label{fig:demo}
\end{figure}

\textbf{Background.}
In a knowledge graph (KG), nodes are entities associated with attributes and interconnected with binary relations as edges. Increasingly many KGs have emerged for various domains, some available as Linked Open Data. KGs for a focused domain often have a simple schema, e.g.,~the LinkedIn KG\footnote{\url{https://www.linkedin.com/pulse/machine-learning-linkedin-knowledge-graph-qi-he/}} describes members, companies, and other entities in the professional domain. There are also generic KGs providing encyclopedic knowledge and hence having a rich schema consisting of thousands of types of relations and attributes, such as DBpedia~\cite{DBpedia}. We illustrate a KG in Fig.~\ref{fig:demo} as the running example in this paper. It describes movies, their actors and directors, and awards.

An established data mining task for KG-based applications is \emph{relevance search}~\cite{pra,srw,pathsim,hetesim,ug,fspg,joinsim,relsim,metagraph,metastructure,prep,relsue}. The task is essentially to find entities in a KG that are the most relevant to an input query entity. However, relevance has a broad range of meanings, particularly over a schema-rich KG. For example, given \texttt{Tom Hardy} as the query entity, the user may search for actresses that co-starred with him, or for American directors that collaborated with him. Unfortunately, non-expert users lack the expertise to formally characterize the desired semantics of relevance because the formal query language and the rich schema of the KG are both difficult to learn.

\textbf{Problem.}
To bridge the gap between non-expert users and structured KGs, one practical solution is to request a small number of examples from the user, and then to \emph{learn user-defined relevance from user-provided examples}~\cite{pra,srw,ug,mpprefer,fspg,relsim,metagraph,relsue}. The user can specify an example in the form of an ordered \emph{query-answer entity pair} to illustrate the desired semantics of relevance~\cite{pra,fspg,relsim}. For example, two different users may provide different sets of examples:
\begin{equation}\label{eq:s1s2}
\begin{split}
    S_1 = \{ & \langle\texttt{Dave Chappelle}, ~\texttt{Lady Gaga}\rangle,\\
    & \langle\texttt{Matt Damon}, ~\texttt{Julia Roberts}\rangle\} \,,\\
    S_2 = \{ & \langle\texttt{Dave Chappelle}, ~\texttt{Bradley Cooper}\rangle,\\
    & \langle\texttt{Matt Damon}, ~\texttt{George Clooney}\rangle\} \,.
\end{split}
\end{equation}

The user providing~$S_1$ aims to find entities that are relevant to \texttt{Tom Hardy} \emph{just as how} \texttt{Lady Gaga} is relevant to \texttt{Dave Chappelle} and as how \texttt{Julia Roberts} is relevant to \texttt{Matt Damon}. The underlying semantics could be actors---preferably American actresses---that co-starred with \texttt{Tom Hardy}. In this case, \texttt{Marion Cotillard} and \texttt{Leonardo DiCaprio} are acceptable answers. The user providing~$S_2$ has a different need and is looking for American male directors and/or actors that collaborated with \texttt{Tom Hardy}. Now, \texttt{Leonardo DiCaprio} and \texttt{Christopher Nolan} become good answers.

\textbf{Challenges.}
It has been common to assume that the user-defined relevance can be represented by one or more \emph{meta-paths}~\cite{pra,pathsim,ug,mpprefer,fspg,relsim,prep,relsue}. A meta-path is a sequence of relation types, i.e.,~a path at the schema level. For example, the co-starring relation underlying~$S_1$ is characterized by the following meta-path:
\begin{equation}\label{eq:mp1}
  \mathcal{P}_1: ~ \texttt{[query]} \xleftarrow{\texttt{stars}} \cdot \xrightarrow{\texttt{stars}} \texttt{[answer]} \,.
\end{equation}
\noindent The collaboration relation underlying~$S_2$ is characterized partially by~$\mathcal{P}_1$ and partially by
\begin{equation}\label{eq:mp2}
    \mathcal{P}_2: ~ \texttt{[query]} \xleftarrow{\texttt{stars}} \cdot \xrightarrow{\texttt{director}} \texttt{[answer]} \,.
\end{equation}

The key challenge to these methods is the selection and weighting of meta-paths. Existing methods train discriminative models to directly learn the weights of meta-paths from user-provided examples~\cite{pra,ug,mpprefer,fspg,relsim,relsue}, but have exhibited the following limitations.
\begin{itemize}
    \item Their \emph{discriminative models} rely on negative examples that are sampled automatically as the user only provides positive examples. Identifying high-quality negative examples is algorithmically challenging and computationally demanding.
    \item They are focused on meta-path based relevance, but cannot support the representation of \emph{properties}, such as gender and nationality in the above example. Their capability of characterizing user needs is somewhat limited.
\end{itemize}

\textbf{Contributions.}
In this work we address the two challenges and propose GREASE, a novel \textbf{G}enerative model for \textbf{R}el\textbf{E}v\textbf{A}nce \textbf{SE}arch over KGs. Our implementation has been open source.\footnote{\url{http://ws.nju.edu.cn/relevance/grease/}} Our contributions are summarized as follows.
\begin{itemize}
    \item We treat the weight of a meta-path as a posterior probability, and devise a \emph{generative model} with cost-effective approximations. Our model does not rely on negative examples, and has outperformed existing methods in both effectiveness and efficiency in the experiments.
    \item We extend the model to support not only meta-path based relevance of answer entities to the query entity, but also \emph{properties that constrain answer entities}. The extension allows to represent more expressive user-defined relevance.
\end{itemize}

\textbf{Organization.}
We formulate the problem in Section~\ref{sect:prob}. Our generative model is described in Section~\ref{sect:app1}, and is extended to support properties in Section~\ref{sect:app2}. Based on that, our search algorithm is given in Section~\ref{sect:app3}. We report experiments in Section~\ref{sect:exp}, compare related work in Section~\ref{sect:rw}, and finally conclude the paper in Section~\ref{sect:con}.

%% file: problem.tex
\section{Problem Formulation}\label{sect:prob}


Let~$\mathbb{R}$ denote the set of all real numbers. We assume countable pairwise disjoint sets of entities~$\mathcal{V}$, relation types~$\mathcal{R}$, attribute types~$\mathcal{A}$, and attribute values~$\mathcal{L}$.

\begin{definition}[Knowledge Graph]\label{def:kg}
	A knowledge graph (KG) is a directed graph denoted by $G= \langle V,E,\Psi \rangle$, where
	\begin{itemize}
		\item $V \subseteq \mathcal{V}$~is a finite set of entities represented as nodes,
		\item $E \subseteq V \times \mathcal{R} \times V$~is a finite set of relations between entities represented as directed edges, and
		\item $\Psi : V \mapsto \powerset(\mathcal{A} \times \mathcal{L})$, where $\powerset(\cdot)$ represents power set, is a function that associates each entity $v \in V$ with a finite set of attributes $\Psi(v) \subseteq \mathcal{A} \times \mathcal{L}$.
	\end{itemize}
\end{definition}


For an entity $v \in V$, its \emph{properties} consist of
\begin{equation}
    \Phi(v) = \{\langle a,l \rangle : \langle a,l \rangle \in \Psi(v) \text{ or } \langle v,a,l \rangle \in E \} \,.
\end{equation}

For example, in Fig.~\ref{fig:demo}, the properties of \texttt{Suburbicon} consist of an attribute $\langle \texttt{genre}, ~\texttt{Comedy} \rangle$ and two relations $\langle \texttt{stars}, ~\texttt{Matt Dameon} \rangle$ and $\langle \texttt{director}, ~\texttt{George Clooney} \rangle$.


We define \emph{path} in a standard way. It is acyclic, and its edges are not required to follow the same direction. However, in the remainder of the paper we always write right arrows and rewrite left arrow $\xleftarrow{r}$ as $\xrightarrow{r^{-1}}$, where $r^{-1}$~represents the inverse of~$r \in \mathcal{R}$.
We write a path $p : v_{0}\xrightarrow{r_{1}}v_{1}\xrightarrow{r_{2}} \cdots \xrightarrow{r_{l}}v_{l}$ from~$v_0$ to~$v_l$ as $v_0 \rightsquigarrow_p v_l$ for short. The length of a path is the number of its edges.

\begin{definition}[Meta-Path]\label{def:mp}
	A meta-path is a sequence of relation types $\mathcal{P}: r_{1} r_{2} \cdots r_{l}$, where $r_1,\ldots,r_l \in \mathcal{R}$ are relation types (or their inverses), and $l$~is called the length of~$\mathcal{P}$ denoted by $\length(\mathcal{P})=l$.
	A~path~$p$ in a KG follows~$\mathcal{P}$, denoted by $p \models \mathcal{P}$, if $p$~is in the form of $p: v_{0}\xrightarrow{r_{1}}v_{1}\xrightarrow{r_{2}} \cdots \xrightarrow{r_{l}}v_{l}$.
\end{definition}

For example, the two meta-paths shown in Eq.~(\ref{eq:mp1}) and Eq.~(\ref{eq:mp2}) are formalized as follows:
\begin{equation}\label{eq:mp12}
    \mathcal{P}_1 : \texttt{stars}^{-1} ~\texttt{stars} \,, \qquad
    \mathcal{P}_2 : \texttt{stars}^{-1} ~\texttt{director} \,.
\end{equation}
\noindent Their lengths are both~2. $\mathcal{P}_2$~is followed by path
\begin{equation}\label{eq:path}
\resizebox{\columnwidth}{!}{
    $\texttt{Tom Hardy} \xrightarrow{\texttt{stars}^{-1}} \texttt{Inception} \xrightarrow{\texttt{director}} \texttt{Christopher Nolan} \,.$
}
\end{equation}



\begin{definition}[Relevance Search]\label{def:rs}
	Given a KG denoted by $G= \langle V,E,\Phi \rangle$, let $\rel : \mathcal{V} \times \mathcal{V} \mapsto \mathbb{R}$ be a user-defined real-valued function. For $u,v \in V$, $\rel(u,v)$ returns the relevance of~$v$ to~$u$. Let~$k < |V|$ be a predetermined positive integer. For an input query entity $q \in V$, relevance search is to find top-$k$ answer entities $\ans(q) \subseteq (V \setminus \{q\})$ that are the most relevant to~$q$ in terms of $\rel$.
\end{definition}


Following~\cite{pra,fspg,relsim}, we assume the user provides a small number of ordered query-answer entity pairs as \emph{examples}, to exemplify the desired $\rel$ which is not directly accessible to the search system.

\begin{definition}[Relevance Search by Example]\label{def:rsupe}
    Extending Definition~\ref{def:rs}, the problem turns into learning a function $\rel$ under the supervision of a set of user-provided examples denoted by~$S$. Each example is an ordered query-answer entity pair $\langle s,t \rangle \in V \times V$, where $s$~is called the source entity and $t$~is the target entity, such that $v \in \ans(q)$ is relevant to~$q$ just as how~$t$ is relevant to~$s$.
\end{definition}

For example, $S_1$~and~$S_2$ in Eq.~(\ref{eq:s1s2}) are two sets of examples for the query entity \texttt{Tom Hardy}. They indicate different $\rel$ functions.

%% file: approach.tex
\section{Generative Relevance Model}\label{sect:app1}

Following Definition~\ref{def:rsupe}, the $\rel$ function is conditioned on a set of user-provided examples~$S$, so we rewrite $\rel(q,v)$ as $\rel(q,v|S)$. Previous research computes meta-path based relevance~\cite{pra,pathsim,ug,mpprefer,fspg,relsim,prep,relsue}. Along this line, we decompose~$\rel$ into a linear combination of weighted relevance over a set of meta-paths denoted by $\boldsymbol{\Omega}_\text{mp}=\{\mathcal{P}_1,\ldots,\mathcal{P}_n\}$:
\begin{equation}\label{eq:rel-mp}
    \rel(q,v|S) = \sum_{\mathcal{P}_i \in \boldsymbol{\Omega}_\text{mp}}{\gamma(q,v|\mathcal{P}_i) \cdot \prb(\mathcal{P}_i|S) \cdot \regularization(\mathcal{P}_i)} \,,
\end{equation}
\noindent where $\gamma(q,v|\mathcal{P}_i)$ measures the real-valued relevance of~$v$ to~$q$ w.r.t. a particular meta-path~$\mathcal{P}_i$, $\prb(\mathcal{P}_i|S)$ represents the weight of~$\mathcal{P}_i$, and $\regularization(\mathcal{P}_i)$ is a regularization term to prevent overfitting. Meta-paths and their weights are to be learned from~$S$. For example, with~$S_1$ in Eq.~(\ref{eq:s1s2}), the meta-path~$\mathcal{P}_1$ in Eq.~(\ref{eq:mp12}) should have a large weight because for every example in~$S_1$, there exists a path in Fig.~\ref{fig:demo} that follows~$\mathcal{P}_1$ and connects the source entity to the target entity.

To establish $\rel$, below we describe the selection of~$\boldsymbol{\Omega}_\text{mp}$, and the computation of~$\gamma$, $\prb$, and~$\regularization$.

\subsection{Meta-Path Selection}

As $\rel$ is exemplified by~$S$, we select~$\boldsymbol{\Omega}_\text{mp}$ based on~$S$. Our~$\boldsymbol{\Omega}_\text{mp}$ contains all possible meta-paths that can be derived from~$S$. A meta-path~$\mathcal{P}_i$ is in~$\boldsymbol{\Omega}_\text{mp}$ if there exists a path in the KG such that it follows~$\mathcal{P}_i$ and it connects the source entity to the target entity in some user-provided example:
\begin{equation}\label{eq:omegamp}
    \boldsymbol{\Omega}_\text{mp} = \bigcup_{\langle s,t \rangle \in S}{\{\mathcal{P} : \exists p \models \mathcal{P}, ~s \rightsquigarrow_p t\}} \,.
\end{equation}

\subsection{Meta-Path Based Relevance}

Extensive research has been conducted to measure the relevance of~$v$ to~$q$ w.r.t. a particular meta-path~$\mathcal{P}_i$~\cite{pc,pra,pathsim,joinsim,hetesim}. This is outside the focus of this paper, and we extend \emph{path count}~\cite{pc} as our measure:
\begin{equation}\label{eq:gammamp}
\begin{split}
    \gamma(q,v|\mathcal{P}_i) & = \min\{\pc(q,v,\mathcal{P}_i), ~\alpha_\text{mp}\} \,,\\
    \pc(q,v,\mathcal{P}_i) & = |\{p : p \models \mathcal{P}_i \text{ and } q \rightsquigarrow_p v\}| \,,
\end{split}
\end{equation}
\noindent where $\pc(q,v,\mathcal{P}_i)$ represents the number of paths in the KG that follow~$\mathcal{P}_i$ and connect~$q$ to~$v$, and $\alpha_\text{mp}>0$ is a parameter to limit the value of~$\pc$ and prevent highly skewed values. For example, in Fig.~\ref{fig:demo}, for~$\mathcal{P}_2$ in Eq.~(\ref{eq:mp12}) we have
\begin{equation}
    \pc(\texttt{Tom Hardy}, ~\texttt{Christopher Nolan}, ~\mathcal{P}_2) = 1 \,,
\end{equation}
\noindent because there is only one path shown in Eq.~(\ref{eq:path}) that follows~$\mathcal{P}_2$ and connects \texttt{Tom Hardy} to \texttt{Christopher Nolan}.

\subsection{Generative Meta-Path Weighting}\label{sect:weightmp}

Weighting scheme is the key to the effectiveness of the $\rel$ function, and is the focus of our work. Different from existing discriminative methods~\cite{pra,ug,mpprefer,fspg,relsim,relsue}, we treat weight~$\prb(\mathcal{P}_i|S)$ as a posterior probability and propose a novel \emph{generative model}. In the following, we will also use~$\prb(\cdot)$ to denote the probability of an event. We estimate probabilities based on the KG and learn~$\prb(\mathcal{P}_i|S)$ from~$S$.

Specifically, we rewrite $\prb(\mathcal{P}_i|S)$ using Bayes' theorem:
\begin{equation}\label{eq:weightmp}
\begin{split}
  \prb(\mathcal{P}_i|S) & = \frac{\prb(\mathcal{P}_i) \cdot \prb(S|\mathcal{P}_i)}{\prb(S)}  \propto \prb(\mathcal{P}_i) \cdot \prb(S|\mathcal{P}_i) \,,
\end{split}
\end{equation}
\noindent where the posterior~$\prb(\mathcal{P}_i|S)$ is proportional to the prior~$\prb(\mathcal{P}_i)$ times the likelihood~$\prb(S|\mathcal{P}_i)$. Below we separately compute the prior and the likelihood.

\textbf{Computation of the Prior.}
For the prior~$\prb(\mathcal{P}_i)$, recall that a meta-path is a sequence of relation types $\mathcal{P}_i : r_1 r_2 \cdots r_l$. We assume the probability of observing the $i$-th relation type~$r_i$ in the context history of the preceding $(i-1)$~relation types $r_1 r_2 \cdots r_{i-1}$ can be approximated by the probability of observing it in the shortened context history of the preceding relation type~$r_{i-1}$, i.e.,~the \emph{first-order Markov property}. This assumption is reasonable and also common on graphs. Specifically, random walks on graphs satisfy this property. Formally, we have
\begin{equation}
\begin{split}
    \prb(\mathcal{P}_i) & = \prb(r_1 r_2 \cdots r_l)  = \prb(r_1) \prod_{i=2}^{l}{\prb(r_i|r_1 r_2 \cdots r_{i-1})} \\
    & \approx \prb(r_1) \prod_{i=2}^{l}{\prb(r_i|r_{i-1})} \,,
\end{split}
\end{equation}
\noindent which in turn will give rise to the following estimation of $\prb(\mathcal{P}_i)$ if we estimate~$\prb(r_1)$ and~$\prb(r_i|r_{i-1})$ from frequency counts:
\begin{equation}\label{eq:priormp1}
    \prb(\mathcal{P}_i) \propto \pc(r_1) \prod_{i=2}^{l}{\frac{\pc(r_{i-1} r_i)}{\pc(r_{i-1})}} \,,
\end{equation}
\noindent where $\pc(\mathcal{P})$~represents the number of paths in the KG that follow meta-path~$\mathcal{P}$:
\begin{equation}\label{eq:pc}
    \pc(\mathcal{P}) = |\{p : p \models \mathcal{P}\}| \,.
\end{equation}

In Eq.~(\ref{eq:priormp1}), computing~$\pc$ according to Eq.~(\ref{eq:pc}) is inexpensive because those meta-paths are very short, being not longer than~2. For example, in Fig.~\ref{fig:demo}, for~$\mathcal{P}_1$ and~$\mathcal{P}_2$ in Eq.~(\ref{eq:mp12}) we have
\begin{equation}
    \pc(\mathcal{P}_1) = 18 \,,\quad \pc(\mathcal{P}_2) = 7 \,.
\end{equation}

Meanwhile, as the reverse direction of~$\mathcal{P}_i$ is also meaningful, we can use the probability of observing the $i$-th relation type~$r_i$ in the context history of the succeeding relation type~$r_{i+1}$, and then make assumptions and estimate probabilities in a similar way:
\begin{equation}\label{eq:priormp2}
\begin{split}
    \prb(\mathcal{P}_i)  \propto \pc(r_l) \prod_{i=1}^{l-1}{\frac{\pc(r_i r_{i+1})}{\pc(r_{i+1})}} \,.
\end{split}
\end{equation}

To improve the robustness of our model, we take the arithmetic mean of Eq.~(\ref{eq:priormp1}) and Eq.~(\ref{eq:priormp2}) as the final value of~$\prb(\mathcal{P}_i)$.

This arithmetic mean also provides an approximation of~$\pc(\mathcal{P})$ for a long meta-path~$\mathcal{P}$. The approximation is useful because it may be infeasible to compute the exact frequency count in Eq.~(\ref{eq:pc}) on a large KG when $\mathcal{P}$~is long. Formally, for $\mathcal{P} : r_1 r_2 \cdots r_l$, let~$\apc(\mathcal{P})$ denote this approximation of $\pc(\mathcal{P})$. We have
\begin{equation}\label{eq:apc}
\begin{split}
    \apc(\mathcal{P}) & =
    \begin{cases}
      \pc(\mathcal{P}) & \length(\mathcal{P}) \leq 2 \,, \\
      \frac{1}{2}(\apc_\text{start}(\mathcal{P}) + \apc_\text{end}(\mathcal{P})) & \length(\mathcal{P}) > 2 \,,
    \end{cases}
\end{split}
\end{equation}
\noindent where $\apc_\text{start}$ and $\apc_\text{end}$ denote the right-hand side of Eq.~(\ref{eq:priormp1}) and Eq.~(\ref{eq:priormp2}), respectively. This approximation will be used later.

\textbf{Computation of the Likelihood.}
For the likelihood~$\prb(S|\mathcal{P}_i)$, the user-provided examples in~$S$ are trivially considered to be conditionally independent given~$\mathcal{P}_i$:
\begin{equation}\label{eq:likelihoodmp}
    \prb(S|\mathcal{P}_i) = \prod_{\langle s,t \rangle \in S}{\prb(\langle s,t \rangle|\mathcal{P}_i)} \,.
\end{equation}

$\prb(\langle s,t \rangle|\mathcal{P}_i)$~represents the probability that a path following~$\mathcal{P}_i$ connects~$s$ to~$t$, which we estimate from frequency counts:
\begin{equation}
    \prb(\langle s,t \rangle|\mathcal{P}_i) \approx \frac{\pc(s,t,\mathcal{P}_i)}{\apc(\mathcal{P}_i)} \,,
\end{equation}
\noindent where $\pc(s,t,\mathcal{P}_i)$ is computed by Eq.~(\ref{eq:gammamp}), and $\apc(\mathcal{P}_i)$ is an approximation of $\pc(\mathcal{P}_i)$ computed by Eq.~(\ref{eq:apc}).

To improve the robustness of our model, \emph{smoothing} is needed to avoid cases where $\pc(s,t,\mathcal{P}_i)=0$ which in turn will lead to $\prb(S|\mathcal{P}_i)=0$. In such a case, we replace the zero value of~$\pc(s,t,\mathcal{P}_i)$ with the following small non-zero value:
\begin{equation}\label{eq:sametype}
  \frac{\apc(\mathcal{P}_i)}{|\sametype(s)| \cdot |\sametype(t)|} \,,
\end{equation}
\noindent where $\apc(\mathcal{P}_i)$ is computed by Eq.~(\ref{eq:apc}), and $\sametype(\cdot)$ denotes the set of entities that have the same (most specific) type as a given entity. Type is an attribute appearing in almost every KG. The above value represents the average number of paths that follow~$\mathcal{P}_i$ and connect two entities of the same type as~$s$ and as~$t$.

\subsection{Regularization}

Long meta-paths are complex and may \emph{overfit user-provided examples}. We impose a penalty on the complexity of~$\mathcal{P}_i$. Specifically, we penalize long meta-paths with the following regularization term:
\begin{equation}\label{eq:beta}
    \regularization(\mathcal{P}_i) = e^{-\beta \cdot \length(\mathcal{P}_i)} \,,
\end{equation}
\noindent where $\length(\mathcal{P}_i)$~denotes the length of~$\mathcal{P}_i$, and $\beta>0$ is a decay factor.

\section{Extended Facet-based Relevance}\label{sect:app2}

Previous research only computes meta-path based relevance~\cite{pra,pathsim,ug,mpprefer,fspg,relsim,prep,relsue}. We extend meta-paths to facets. A \emph{facet} is either a meta-path or a property that constrains answer entities. Accordingly, we extend the~$\rel$ function in Eq.~(\ref{eq:rel-mp}) by adding a linear combination of weighted relevance over a set of properties denoted by $\boldsymbol{\Omega}_\text{prop} = \{\langle a_1,l_1 \rangle,\ldots,\langle a_n,l_n \rangle\}$:
\begin{equation}\label{eq:rel}
\begin{split}
    \rel(q,v|S) & = \sum_{\mathcal{P}_i \in \boldsymbol{\Omega}_\text{mp}}{\gamma(q,v|\mathcal{P}_i) \cdot \prb(\mathcal{P}_i|S) \cdot \regularization(\mathcal{P}_i)} \\
    & +  \sum_{\langle a_i,l_i \rangle \in \boldsymbol{\Omega}_\text{prop}}{\gamma(q,v|\langle a_i,l_i \rangle) \cdot \prb(\langle a_i,l_i \rangle|S)} \,,
\end{split}
\end{equation}
\noindent where $\gamma(q,v|\langle a_i,l_i \rangle)$ measures the real-valued relevance of~$v$ to~$q$ w.r.t. a particular property~$\langle a_i,l_i \rangle$, and $\prb(\langle a_i,l_i \rangle|S)$ represents the weight of~$\langle a_i,l_i \rangle$. A property is already simple and hence regularization is not needed. Properties and their weights are also to be learned from~$S$. For example, with~$S_1$ in Eq.~(\ref{eq:s1s2}), two properties $\langle \texttt{gender},~\texttt{F} \rangle$ and $\langle \texttt{country},~\texttt{US} \rangle$ should have large weights because they constrain all the target entities in~$S_1$.

To establish the extended $\rel$, below we describe the selection of~$\boldsymbol{\Omega}_\text{prop}$, and the computation of~$\gamma$ and $\prb$ for properties.

\subsection{Property Selection}

A property~$\langle a_i,l_i \rangle$ is in~$\boldsymbol{\Omega}_\text{prop}$ if it constrains the target entity in some user-provided example:
\begin{equation}\label{eq:omegaprop}
    \boldsymbol{\Omega}_\text{prop} = \bigcup_{\langle s,t \rangle \in S}{\Phi(t)} \,.
\end{equation}

\subsection{Property-Based Relevance}

We measure the relevance of~$v$ w.r.t.~$\langle a_i,l_i \rangle$ according to whether $\langle a_i,l_i \rangle$ is a property that constrains~$v$, which is independent of~$q$:
\begin{equation}\label{eq:gammaprop}
  \gamma(q,v|\langle a_i,l_i \rangle) = 
  \begin{cases}
    \alpha_\text{prop} & \langle a_i,l_i \rangle \in \Phi(v) \,,\\
    0 & \langle a_i,l_i \rangle \notin \Phi(v) \,,
  \end{cases}
\end{equation}
\noindent where $\alpha_\text{prop}>0$ is a parameter to tune the importance of properties relative to meta-paths in the computation of the extended $\rel$.

\subsection{Generative Property Weighting}\label{sect:weightprop}

Similar to our generative model for weighting meta-paths, we treat weight~$\prb(\langle a_i,l_i \rangle|S)$ as a posterior probability, and rewrite it using Bayes' theorem:
\begin{equation}\label{eq:weightprop}
    \prb(\langle a_i,l_i \rangle|S) \propto \prb(\langle a_i,l_i \rangle) \cdot \prb(S|\langle a_i,l_i \rangle) \,,
\end{equation}
\noindent where the posterior~$\prb(\langle a_i,l_i \rangle|S)$ is proportional to the prior~$\prb(\langle a_i,l_i \rangle)$ times the likelihood~$\prb(S|\langle a_i,l_i \rangle)$. Below we separately compute the prior and the likelihood.

\textbf{Computation of the Prior.}
For the prior~$\prb(\langle a_i,l_i \rangle)$, it represents the probability that the property $\langle a_i,l_i \rangle$ constrains an entity in the KG, which we estimate from frequency counts:
\begin{equation}\label{eq:priorprop}
    \prb(\langle a_i,l_i \rangle) = \frac{|\{v \in V : \langle a_i,l_i \rangle \in \Phi(v)\}|}{|V|} \,,
\end{equation}
\noindent where $V$~is the set of entities in the KG.

\textbf{Computation of the Likelihood.}
For the likelihood~$\prb(S|\langle a_i,l_i \rangle)$, similar to Eq.~(\ref{eq:likelihoodmp}), the user-provided examples in~$S$ are trivially considered to be conditionally independent given~$\langle a_i,l_i \rangle$:
\begin{equation}\label{eq:likelihoodprop}
\begin{split}
    \prb(S|\langle a_i,l_i \rangle) & = \prod_{\langle s,t \rangle \in S}{\prb(\langle s,t \rangle | \langle a_i,l_i \rangle)}  = \prod_{\langle s,t \rangle \in S}{\prb(t | \langle a_i,l_i \rangle)} \,,
\end{split}
\end{equation}
\noindent where the last equation holds because the property~$\langle a_i,l_i \rangle$ constrains answer entities which correspond to the target entity~$t$ in a user-provided example, and hence $\langle a_i,l_i \rangle$~is independent of the source entity~$s$ in the example.

$\prb(t | \langle a_i,l_i \rangle)$~represents the probability that an entity constrained by~$\langle a_i,l_i \rangle$ is~$t$, which we estimate from frequency counts:
\begin{equation}\label{eq:likelihoodpropt}
    \prb(t|\langle a_i,l_i \rangle) =
    \begin{cases}
      \frac{1}{|\{v \in V : \langle a_i,l_i \rangle \in \Phi(v)\}|} & \langle a_i,l_i \rangle \in \Phi(t) \,,\\
      0 & \langle a_i,l_i \rangle \notin \Phi(t) \,.
    \end{cases}
\end{equation}

To improve the robustness of our model, \emph{smoothing} is also needed here to avoid cases where $\prb(t|\langle a_i,l_i \rangle)=0$ which in turn will lead to $\prb(S|\langle a_i,l_i \rangle)=0$. In such a case, we replace the zero value of~$\prb(t|\langle a_i,l_i \rangle)$ with a small non-zero value~$\frac{1}{|V|}$.

\section{Search Algorithm}\label{sect:app3}

Finally, we present an efficient implementation to support learning the proposed model from user-provided examples in an online environment and returning answer entities promptly.

\begin{figure}[t]
\begin{algorithmic}[1]
  \REQUIRE A KG $G=\langle V,E,\Psi \rangle$, a query entity~$q$, a set of user-provided examples~$S$, an upper bound~$L$ on the length of allowable meta-paths, and a positive integer~$m$.
  \ENSURE $k$~top-ranked entities that are relevant to~$q$.
  \STATE $\boldsymbol{\Omega}_\text{mp} \leftarrow$ MPSearch($G,S,L$);
  \STATE $\boldsymbol{\Omega}_\text{prop} \leftarrow \bigcup_{\langle s,t \rangle \in S}{\Phi(t)}$;
  \STATE $\boldsymbol{\Omega} \leftarrow \boldsymbol{\Omega}_\text{mp} \cup \boldsymbol{\Omega}_\text{prop}$;
  \FORALL{$\Omega_i \in \boldsymbol{\Omega}$}
    \STATE Compute $\prb(\Omega_i|S)$;
  \ENDFOR
  \STATE $\boldsymbol{\Omega}_\text{top} \leftarrow$ $m$~meta-paths in~$\boldsymbol{\Omega}_\text{mp}$ with the largest weights;
  \STATE $C \leftarrow \bigcup_{\mathcal{P}_i \in \boldsymbol{\Omega}_\text{top}}{\{v \in V : \exists p \models \mathcal{P}_i, ~q \rightsquigarrow_p v\}}$;
  \FORALL{$v \in C$}
    \STATE Compute $\rel(q,v|S)$;
  \ENDFOR
  \RETURN $k$~top-ranked entities in~$C$
\end{algorithmic}
\caption{The GREASE algorithm.}
\label{alg:alg}
\end{figure}

\subsection{Algorithm}

The full GREASE algorithm is presented in Fig.~\ref{alg:alg}. MPSearch (line~1) finds~$\boldsymbol{\Omega}_\text{mp}$ which is defined by Eq.~(\ref{eq:omegamp}). It performs $|S|$~bidirectional searches---one for each example in~$S$, starting simultaneously from the source entity and the target entity. The search space is restricted to meta-paths that are not longer than~$L$, which is a predetermined upper bound. Then we find~$\boldsymbol{\Omega}_\text{prop}$ (line~2) which is defined by Eq.~(\ref{eq:omegaprop}). $\boldsymbol{\Omega}_\text{mp}$~and~$\boldsymbol{\Omega}_\text{prop}$ comprise~$\boldsymbol{\Omega}$ (line~3), namely all the facets to consider for computing~$\rel$ in Eq.~(\ref{eq:rel}). For each facet $\Omega_i \in \boldsymbol{\Omega}$, we compute~$\prb(\Omega_i|S)$ according to Section~\ref{sect:weightmp} and Section~\ref{sect:weightprop} (lines~4--6). The $m$~meta-paths in~$\boldsymbol{\Omega}_\text{mp}$ with the largest weights are denoted by~$\boldsymbol{\Omega}_\text{top}$ (line~7). All the entities that are connected from~$q$ by a path following a meta-path in~$\boldsymbol{\Omega}_\text{top}$ form candidate answer entities, denoted by~$C$ (line~8). Here, we only use $m$~meta-paths to identify candidate answer entities for efficiency considerations. For each candidate answer entity, its extended relevance to~$q$ is computed (lines~9--11), and the $k$~top-ranked ones are returned.

\subsection{Indexing}

To support efficient computation in an online environment, we precompute and index the following statistics.

We index the frequency counts for all the meta-paths in the KG that are not longer than~2, so their $\pc$~values defined by Eq.~(\ref{eq:pc}) are retrievable in $O(1)$, and for any other meta-path, its approximate $\pc$~value (i.e.,~$\apc$) defined by Eq.~(\ref{eq:apc}) is computable in $O(L)$.

We also index the frequency counts for all the properties in the KG, so Eq.~(\ref{eq:priorprop}) and Eq.~(\ref{eq:likelihoodpropt}) are computable in $O(1)$.

We index the frequency counts for all the entity types in the KG, so $|\sametype(\cdot)|$ in Eq.~(\ref{eq:sametype}) is computable in $O(1)$.

\subsection{Complexity Analysis}

Let~$\Delta$ be the maximum degree of the nodes in the KG. Let~$\Xi$ be the maximum number of properties that constrain an entity in the KG. $\boldsymbol{\Omega}_\text{mp}$~is computed in $O(|S| \cdot \Delta^{\lceil \frac{L}{2} \rceil})$ using bidirectional search, and $\boldsymbol{\Omega}_\text{prop}$~is computed in $O(|S| \cdot \Xi)$.

To compute the posterior~$\prb(\Omega_i|S)$, when $\Omega_i$~is a meta-path~$\mathcal{P}_i$, the prior~$\prb(\mathcal{P}_i)$ is computed by Eq.~(\ref{eq:priormp1}) and Eq.~(\ref{eq:priormp2}) in~$O(L)$, and the likelihood~$\prb(S|\mathcal{P}_i)$ is computed by Eq.~(\ref{eq:likelihoodmp})  in~$O(L+|S|)$ where $\pc(s,t,\mathcal{P}_i)$ has been computed during MPSearch. When $\Omega_i$~is a property~$\langle a_i,l_i \rangle$, the prior~$\prb(\langle a_i,l_i \rangle)$ is computed by Eq.~(\ref{eq:priorprop}) in~$O(1)$, and the likelihood~$\prb(S|\langle a_i,l_i \rangle)$ is computed by Eq.~(\ref{eq:likelihoodprop}) in~$O(|S|)$.

The candidate answer entities~$C$ are computed in~$O(m \cdot \Delta^L)$.

The computation of the $\rel$ function does not further increase the asymptotic time complexity of the algorithm.

To conclude, in practice, the running time of the entire algorithm is probably dominated by MPSearch.

%% file: experiment.tex
\section{Experiments}\label{sect:exp}
We empirically compared our approach with several state-of-the-art methods based on a variety of queries over two popular KGs. Both effectiveness and efficiency were tested.

\subsection{Datasets}

\begin{table}[t]
\caption{Statistics about KGs}
\begin{center}
\resizebox{\columnwidth}{!}{
\begin{tabular}{|l|r|r|r|r|} 
\hline
\multirow{2}{*}{\textbf{KG}}& \multirow{2}{*}{\textbf{Entity}} & \multirow{2}{*}{\textbf{Relation}} & \textbf{Relation} & \textbf{Attribute} \\
& & & \textbf{Type} & \textbf{Type} \\
\hline
DBpedia 2016-10 & \multirow{2}{*}{5,900,558} & \multirow{2}{*}{18,746,174} & \multirow{2}{*}{661} & \multirow{2}{*}{2,065} \\
(Mappingbased Objects) & & & & \\
\hline
YAGO 3.1 & \multirow{2}{*}{4,295,825} & \multirow{2}{*}{12,430,700} & \multirow{2}{*}{37} & \multirow{2}{*}{1} \\
(yagoFacts) & & & & \\
\hline
\end{tabular}
}
\end{center}
\label{tab:ds}
\end{table}

\textbf{Knowledge Graphs.}
Our experiments were based on two popular large-scale KGs: DBpedia~\cite{DBpedia} (version~2016-10) and YAGO~\cite{YAGO} (version~3.1). For DBpedia, we obtained a KG from two files: \emph{Mappingbased Objects} and \emph{Instance Types}. For YAGO, we obtained a KG from two files: \emph{yagoFacts} and \emph{yagoSimpleTypes}. The files are in RDF format, where RDF literals and types were treated as attributes, and the other RDF triples became relations. Some statistics about these KGs are summarized in Table~\ref{tab:ds}. Note that YAGO was to be used with the queries created in~\cite{relsue} where attributes are not involved. So we followed~\cite{relsue} to only import \emph{yagoFacts} and \emph{yagoSimpleTypes} which contain relations but no attributes except for type.

\begin{table*}[t]
\caption{Query Groups with Examples}
\begin{center}
\resizebox{\textwidth}{!}{
\begin{tabular}{|c|l|c|c|p{5cm}|c|}
\hline
& \multirow{2}{*}{\textbf{Desired Semantics (Facets)}} & \multicolumn{4}{|c|}{\textbf{Example Query and Answers}} \\
\cline{3-6}
& & \textbf{\textit{Query Entity}} & \textbf{\textit{User-Provided Examples}} & \textbf{\textit{Query Intent}} & \textbf{\textit{Answer Entities}} \\
\hline
\multirow{2}{*}{D1} & \multirow{2}{*}{starring$^{-1}$ director} & \multirow{2}{*}{Howard Duff} & $\langle \text{Stephen Wight}, ~\text{Susan Tully} \rangle$ & \multirow{2}{*}{\parbox{5cm}{director of a movie starring Howard Duff}} & George Sherman \\
& & & $\langle \text{Vijay Chavan}, ~\text{Kedar Shinde} \rangle$ & & Andre deToth \\
\hline
\multirow{2}{*}{D2} & \multirow{2}{*}{almaMater$^{-1}$ foundedBy$^{-1}$} & \multirow{2}{*}{Bowdoin College} & $\langle \text{Duke University}, ~\text{Duolingo} \rangle$ & \multirow{2}{*}{\parbox{5cm}{organization founded by a Bowdoin alumnus}} & Netflix \\
& & & $\langle \text{Yale University}, ~\text{Allied Corp} \rangle$ & & Pure Software \\
\hline
\multirow{2}{*}{D3} & starring starring$^{-1}$ & \multirow{2}{*}{Charlie Chaplin} & $\langle \text{Bam Margera}, ~\text{Brandon Novak} \rangle$ & \multirow{2}{*}{\parbox{5cm}{actor and also director of a movie starring Charlie Chaplin}} & Mack Swain \\
& starring director$^{-1}$ & & $\langle \text{Rahul Bose}, ~\text{Koel Purie} \rangle$ & & Lloyd Bacon \\
\hline
\multirow{2}{*}{D4} & almaMater almaMater$^{-1}$ & \multirow{2}{*}{Hagan Bayley} & $\langle \text{Ewan Birney}, ~\text{Antony Galione} \rangle$ & \multirow{2}{*}{\parbox{5cm}{Hagan Bayley's schoolmate that won the same award}} & John Mollon \\
& award award$^{-1}$ & & $\langle \text{George Porter}, ~\text{Charles Coulson} \rangle$ & & Henry Gilman \\
\hline
\multirow{2}{*}{D5} & architecturalStyle architecturalStyle$^{-1}$ & \multirow{2}{*}{De Rohan Arch} & $\langle \text{Morson's Row}, ~\text{PaceKing House} \rangle$ & \multirow{2}{*}{\parbox{5cm}{architecture with the same style and location as the De Rohan Arch}} & Hompesch Gate \\
& location location$^{-1}$ & & $\langle \text{Evergreen (Virginia)}, ~\text{Greer House} \rangle$ & & La Borsa \\
\hline
\multirow{2}{*}{D6} & starring$^{-1}$ director director$^{-1}$  & \multirow{2}{*}{Tanya Chisholm} & $\langle \text{Casey Kasem}, ~\text{Fantastic Max} \rangle$ & \multirow{2}{*}{\parbox{5cm}{comedy directed by the director of a movie starring Tanya Chisholm}} & The Last Halloween \\
& $\langle \text{genre}, ~\text{Comedy} \rangle$ & & $\langle \text{Michael Milhoan}, ~\text{Party Down} \rangle$ & & A Fairly Odd Summer \\
\hline
\multirow{2}{*}{D7} & almaMater almaMater$^{-1}$ foundedBy$^{-1}$  & \multirow{2}{*}{Vitalik Buterin} & $\langle \text{Peter Clyne}, ~\text{SpringSource} \rangle$ & \multirow{2}{*}{\parbox{5cm}{software company founded by Vitalik Buterin's schoolmate}} & Databricks \\
& $\langle \text{industry}, ~\text{Software} \rangle$ & & $\langle \text{Felix Villars}, ~\text{Lightbend Inc.} \rangle$ & & Waterloo Maple \\
\hline
\multirow{2}{*}{D8} & influenced$^{-1}$ influenced & \multirow{2}{*}{Leo Strauss} & $\langle \text{Denis Diderot}, ~\text{Leonhard Euler} \rangle$ & \multirow{2}{*}{\parbox{5cm}{physicist influenced by the same person as Leo Strauss}} & Isaac Newton \\
& $\langle \text{field}, ~\text{Physics} \rangle$ & & $\langle \text{Herbert Feigl}, ~\text{Albert Einstein} \rangle$ & & David Hilbert \\
\hline
\multirow{2}{*}{D9} & affiliation affiliation$^{-1}$ & \multirow{2}{*}{Carleton College} & $\langle \text{Viterbo University}, ~\text{Fisk University} \rangle$ & \multirow{2}{*}{\parbox{5cm}{private school affiliated with the same organization as Carleton College}} & Manhattan College \\
& $\langle \text{type}, ~\text{Private school} \rangle$ & & $\langle \text{Verdon College}, ~\text{DePaul University} \rangle$ & & Drake University \\
\hline
\multirow{2}{*}{D10} & museum$^{-1}$ author & \multirow{2}{*}{Metropolitan Museum of Art} & $\langle \text{National Gallery}, ~\text{Georges Seurat} \rangle$ & \multirow{2}{*}{\parbox{5cm}{Parisian artist with  artworks housed by Metropolitan Museum of Art}} & Georges Seurat \\
& $\langle \text{birthPlace}, ~\text{Paris} \rangle$ & & $\langle \text{Van Gogh Museum}, ~\text{Robert Delaunay} \rangle$ & & Jacques-Louis David \\
\hline
\end{tabular}
}
\end{center}
\label{tab:query}
\end{table*}

\textbf{Queries.}
We reused 320~query instances\footnote{\url{http://ws.nju.edu.cn/relevance/relsue/}} given in~\cite{relsue} which were divided into 8~groups (D11--D14 and Y1--Y4). However, this dataset is still limited in two aspects. First, attributes are not considered. Second, the query entity is required to appear as the source entity in every user-provided example, because the method proposed in~\cite{relsue} was specifically designed for this scenario. To overcome these two limitations,  we created 800~query instances\footnote{\url{http://ws.nju.edu.cn/relevance/grease/}} which were divided into 10~groups (D1--D10).

Our creation of D1--D10 followed a common practice in related work~\cite{relsim,relsue}. Compared with D11--D14 and Y1--Y4 created in~\cite{relsue}, our D1--D10 are more generalized as they allow the source entity in a user-provided example to be different from the query entity, and they are more challenging as D6--D10 involve properties. In general, their desired semantics of relevance are more complex than all the known queries used in the literature.

Specifically, each group of D1--D10 contains 80~query instances, and their desired semantics of relevance are represented by the same set of predefined facets. We sampled 100~random source-target entity pairs from DBpedia as a pool such that their relevance conformed to the predefined semantics. We chose 20~random pairs from the pool and took their source entities as our query entities. For each query entity, based on the predefined semantics of relevance, we labeled gold-standard answer entities, and created 4~query instances by choosing different numbers of random pairs from the pool as user-provided examples: $|S| \in \{2,3,4,5\}$. Table~\ref{tab:query} illustrates each group with one query instance under $|S|=2$ and two of its gold-standard answer entities.

The creation of D11--D14 and Y1--Y4 in~\cite{relsue} adopted a similar procedure, and we refer the reader to~\cite{relsue} for details. D11--D14 are based on DBpedia, and Y1--Y4 are based on YAGO. Each group contains 40~query instances.

\subsection{Baselines}
To compare with the state of the art, we chose five strong baselines: PRA~\cite{pra}, RelSim~\cite{relsim}, RelSUE~\cite{relsue}, ProxE~\cite{proxembed}, and D2AGE~\cite{d2age}. We intended to also compare with FSPG~\cite{fspg}, but we could not obtain its implementation from its authors and we failed to re-implement it due to some missing details in the algorithm.

All the chosen baseline methods except for RelSUE could be tested with all the query instances in our experiments. RelSUE requires the query entity to appear as the source entity in every user-provided example, so it could only be tested with D11--D14 and Y1--Y4 created by the authors of RelSUE.

We obtained implementations of RelSUE, ProxE, and D2AGE from their authors, and we re-implemented PRA and RelSim. For PRA, RelSim, ProxE, and D2AGE, we consistently set their bounds on meta-path length to~3, being sufficiently large for representing all the semantics of relevance in our experiments. RelSUE automatically generated meta-paths of varied lengths.

PRA, RelSim, and RelSUE automatically sampled $10 \cdot |S|$~negative examples for training. For ProxE and D2AGE, a training example is a triple $\langle u,v,w \rangle$ where entity~$v$ is more relevant to query entity~$u$ than entity~$w$. We generated 100~such triples for each query instance by extending our positive example $\langle u,v \rangle$ with a random entity~$w$ having the same type as~$v$.
\subsection{Configuration and Variant of GREASE}
For our proposed approach GREASE, by default we set $\alpha_\text{mp}=5$ in Eq.~(\ref{eq:gammamp}), $\alpha_\text{prop}=2$ in Eq.~(\ref{eq:gammaprop}), $\beta=10$ in Eq.~(\ref{eq:beta}), $L=3$ and $m=3$ in Algorithm~\ref{alg:alg}. A parameter study will be reported in Section~\ref{sect:para}.

For D1--D5, D11--D14, and Y1--Y4 where only meta-paths but no properties are involved, we implemented a variant of GREASE using only meta-paths as facets, denoted by GREASE-np.


\subsection{Effectiveness Evaluation}

\textbf{Evaluation Metric.}
For each query instance, the gold standard is a set of relevant answer entities. Each tested method computed $k$~top-ranked answer entities. Their quality was measured by the Normalized Discounted Cumulative Gain (NDCG) at rank position~$k$, referred to as NDCG@$k$. Due to space limitations, we only present NDCG@10 scores (i.e.,~$k=10$).

\begin{table}[t]
\caption{NDCG@10 on D1--D5}
\begin{center}
\begin{tabular}{|c|c|c|c|c|} 
\hline
Method & $|S|=2$  & $|S|=3$ & $|S|=4$ & $|S|=5$ \\
\hline
PRA & 0.535 & 0.636 & 0.599 & 0.639 \\ 
RelSim & 0.452 & 0.562 & 0.555 & 0.575 \\
ProxE & 0.484 &	0.467 &	0.480 &	0.503 \\ 
D2AGE & 0.549 &	0.509 &	0.555 &	0.637 \\ 
\hline
GREASE & 0.782 & 0.737 & 0.734 & 0.763 \\
GREASE-np & \textbf{0.846} & \textbf{0.850} & \textbf{0.865} & \textbf{0.862} \\ 
\hline
\end{tabular}
\end{center}
\label{tab:d15}
\end{table}

\begin{table}[t]
\caption{NDCG@10 on D6--D10}
\begin{center}
\begin{tabular}{|c|c|c|c|c|} 
\hline
Method & $|S|=2$ & $|S|=3$ & $|S|=4$ & $|S|=5$ \\
\hline
PRA & 0.295 & 0.293 & 0.337 & 0.339 \\ 
RelSim & 0.329 & 0.403 & 0.428 & 0.439 \\ 
ProxE & 0.445 &	0.440 &	0.450 &	0.426 \\ 
D2AGE & 0.366 &	0.381 &	0.397 &	0.401 \\ 
\hline
GREASE & \textbf{0.831} & \textbf{0.840} & \textbf{0.866} & \textbf{0.874} \\ 
\hline
\end{tabular}
\end{center}
\label{tab:d610}
\end{table}

\textbf{Results on D1--D5.}
The average NDCG@10 scores on D1--D5 are presented in Table~\ref{tab:d15}. The results are categorized by number of user-provided examples (i.e.,~$|S|$). GREASE outperformed all the baselines by at least 0.101--0.233 under different values of~$|S|$. GREASE-np achieved even higher scores, exceeding the baselines by at least 0.214--0.297. Recall that for the query instances in D1--D5, their desired semantics of relevance are represented by only meta-paths. Therefore, the superiority of GREASE-np over the baselines demonstrated the effectiveness of our proposed generative model for weighting meta-paths.

Compared with GREASE-np, the small drops in GREASE's scores suggested that its extended model mistakenly assigned large weights to some properties. The extended model is expressive in characterizing user-defined relevance but is then more prone to errors due to the expansion of the search space. Nevertheless, the satisfying performance of GREASE showed that it achieved a good trade-off between expressiveness and accuracy.

Another finding was GREASE and GREASE-np already achieved high scores when $|S|=2$. Their performance did not increase notably when $|S|$~increased. It indicated that using our approach, users can obtain quite accurate answers with a small effort.

\textbf{Results on D6--D10.}
The average NDCG@10 scores on D6--D10 are presented in Table~\ref{tab:d610}. Query instances in D6--D10 require using properties that constrain answer entities. Not surprisingly, GREASE largely surpassed all the baselines by at least 0.386--0.435 under different values of~$|S|$, as it was the only method that explicitly processed properties. It confirmed the expressiveness of our extended model in supporting the representation of user-defined relevance.

\begin{table}[t]
\caption{NDCG@10 on D11--D14}
\begin{center}
\begin{tabular}{|c|c|c|c|c|} 
\hline
Method & $|S|=2$ & $|S|=3$ & $|S|=4$ & $|S|=5$ \\ 
\hline
PRA & 0.465 & 0.520 & 0.550 & 0.568 \\
RelSim & 0.644 & 0.656 & 0.654 & 0.666 \\ 
RelSUE & 0.901 & 0.952 & 0.948 & 0.971 \\ 
ProxE & 0.410 &	0.402 &	0.410 &	0.371 \\ 
D2AGE & 0.627 & 0.672 &	0.746 &	0.697 \\
\hline
GREASE & 0.971 & \textbf{0.978} & \textbf{0.953} & \textbf{0.973} \\
GREASE-np & \textbf{0.995} & 0.968 & 0.942 & 0.968 \\
\hline
\end{tabular}
\end{center}
\label{tab:d1114}
\end{table}

\begin{table}[t]
\caption{NDCG@10 on Y1--Y4}
\begin{center}
\begin{tabular}{|c|c|c|c|c|}
\hline
Method & $|S|=2$ & $|S|=3$ & $|S|=4$ & $|S|=5$ \\
\hline
PRA & 0.215 & 0.144 & 0.144 & 0.181 \\
RelSim & 0.274 & 0.336 & 0.357 & 0.367 \\
RelSUE & \textbf{0.770} & 0.843 & \textbf{0.873} &	0.880 \\
ProxE & 0.568 &	0.592 &	0.562 &	0.608 \\
D2AGE & 0.670 &	0.637 &	0.735 &	0.647 \\
\hline
GREASE & 0.724 & \textbf{0.861}	& 0.860 & \textbf{0.900} \\
GREASE-np & 0.673 &	0.677 &	0.674 &	0.703 \\
\hline
\end{tabular}
\end{center}
\label{tab:y14}
\end{table}

\textbf{Results on D11--D14 and Y1--Y4.}
The average NDCG@10 scores on D11--D14 and Y1--Y4 are presented in Table~\ref{tab:d1114} and Table~\ref{tab:y14}, respectively. Query instances in D11--D14 and Y1--Y4 represent a special case of our problem, where the query entity appears as the source entity in every user-provided example. RelSUE was specifically optimized for this scenario and represented the state of the art. Its scores were very high on D11--D14, in the range of 0.901--0.972 under different values of~$|S|$. GREASE performed even better, although their differences were not large: 0.002--0.070. On Y1--Y4, GREASE led when $|S|=3$ and $|S|=5$, whereas RelSUE was better when $|S|=2$ and $|S|=4$. We concluded that GREASE was comparable with RelSUE in this special setting. It demonstrated the effectiveness and generalizability of our approach.


\begin{figure}[t]
	\begin{subfigure}{\linewidth}
	 \includegraphics[width=\columnwidth]{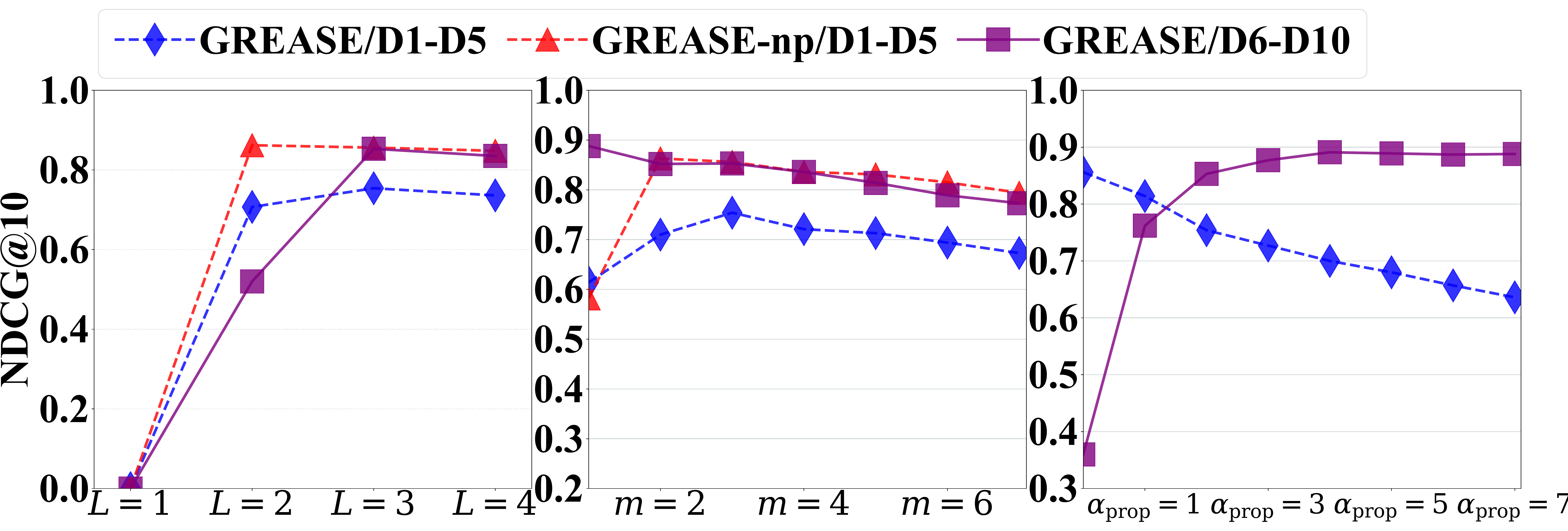}
	 \label{fig:para_1}
	\end{subfigure}
	\begin{subfigure}{\linewidth}
	 \includegraphics[width=\columnwidth]{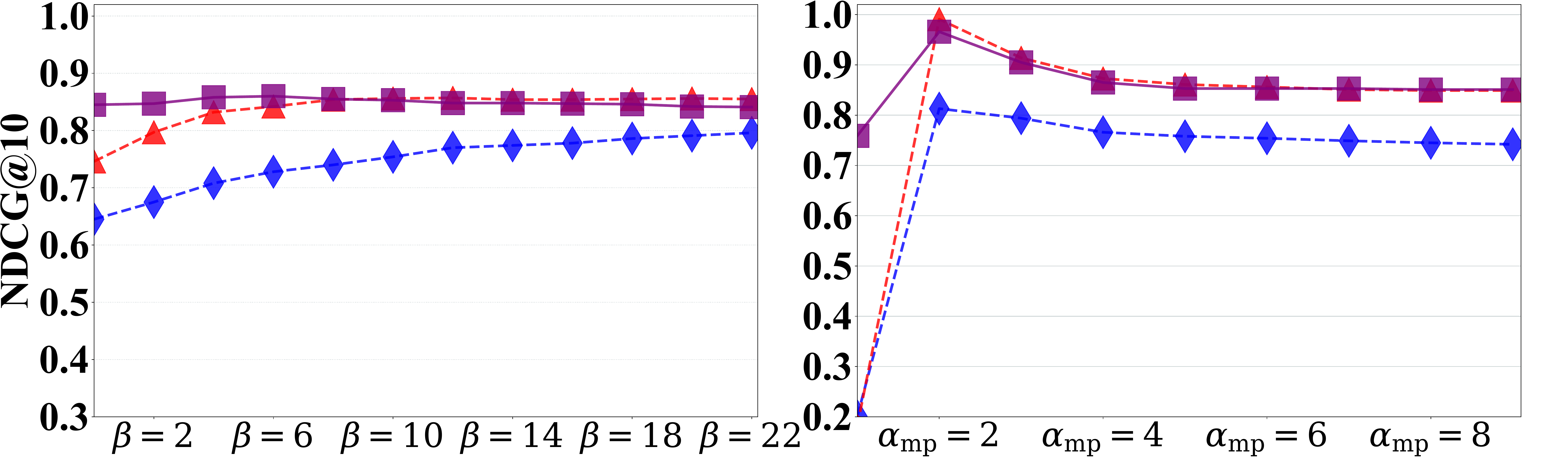}
	 \label{fig:para_2}
	\end{subfigure}
	\caption{Influence of parameters on our approach.}
	\label{fig:para}
\end{figure}

\subsection{Parameter Study}\label{sect:para}

In Fig.~\ref{fig:para}, we present the influence of the five parameters on our approach: $L$, $m$, $\alpha_\text{prop}$, $\beta$, and~$\alpha_\text{mp}$.

$L$~and~$m$ in Algorithm~\ref{alg:alg} tune the trade-off between expressiveness and efficiency in our approach. $L$~bounds the length of allowable meta-paths, and $m$~bounds the number of meta-paths used to identify candidate answer entities. All the desired semantics of relevance in our experiments can be represented by meta-paths not longer than~3. GREASE exactly peaked when $L=3$. Its scores did not change much when $L$~increased to~4. That would give us more flexibility in practice. As to~$m$, our approach achieved good results when $m$~was small, owing to the high quality of the computed meta-paths with the largest weights. When $m$~was larger, more noise could be introduced, but the performance of our approach appeared rather stable.

$\alpha_\text{prop}$~in Eq.~(\ref{eq:gammaprop}) tunes the importance of properties relative to meta-paths. On D1--D5 where properties are not needed, larger values of~$\alpha_\text{prop}$ led to poorer results. The setting of this parameter would depend on the needs of the application.

$\beta$~in Eq.~(\ref{eq:beta}) tunes the degree of penalizing long meta-paths to prevent overfitting. $\alpha_\text{mp}$~in Eq.~(\ref{eq:gammamp}) bounds the value of path count to prevent highly skewed values. The performance of GREASE was more sensitive to~$\beta$ on D1--D5 than on D6--D10, because D1--D5 totally rely on meta-paths. The performance was generally not very sensitive to~$\alpha_\text{mp}$ unless it was inappropriately set to~1 which disabled path count.

\subsection{Efficiency Evaluation}

Our experiments were performed on a 3.40GHz Xeon. 
The precomputed indexes for GREASE only used 190MB for DBpedia and 123MB for YAGO.

In Fig.~\ref{fig:timeall}, for each method we report its average running time per query instance. GREASE satisfyingly completed a search task in less than~1s on DBpedia (D1--D14), at least an order of magnitude faster than all the baselines. It used less than~10s on YAGO (Y1--Y4), being comparable with RelSUE which was optimized for this special scenario. The results demonstrated the efficiency of our approach.

PRA, RelSim, and GREASE ran slower on YAGO than on DBpedia.
These methods search the KG to generate all possible meta-paths of a bounded length. YAGO contains much more paths to explore than DBpedia---about 40~times in our experiments, due to the existence of hub nodes in YAGO.

In Fig.~\ref{fig:timeour}, we show the influence of~$|S|$ on the efficiency of GREASE. The influence was not significant on DBpedia (D1--D14), which suggested the scalability of our approach. However, on YAGO (Y1--Y4), the running time exhibited a linear correlation with~$|S|$.

%% file: related_work_icdm.tex
\section{Related Work}\label{sect:rw}

\subsection{Unsupervised Similarity Search}

Relevance search originates from similarity search, for which methods are mainly based on random walks.
ObjectRank~\cite{objectrank} computes the stationary probability that a random surfer starting from the query entity is at a particular entity as their similarity.
PathSim~\cite{pathsim} requires random walks to follow a predefined meta-path, to compute similarity with a specified type of semantics.
JoinSim~\cite{joinsim} uses a slightly different measure.
PReP~\cite{prep} extends PathSim and JoinSim. It computes cross-meta-path synergy, which goes beyond a linear combination of meta-paths.
In~\cite{metastructure}, meta-path is extended to meta-structure which is a directed acyclic graph.

These methods have found application in entity resolution and entity clustering~\cite{PathSelClus} where similarity measurement is the core task.
However, similarity is only one special type of relevance. Without the supervision of the user, these methods are not suitable for the more generalized relevance search because they cannot distinguish between a wide range of semantics of relevance on a KG.

\begin{figure}[t]
\centerline{\includegraphics[width=0.8\columnwidth]{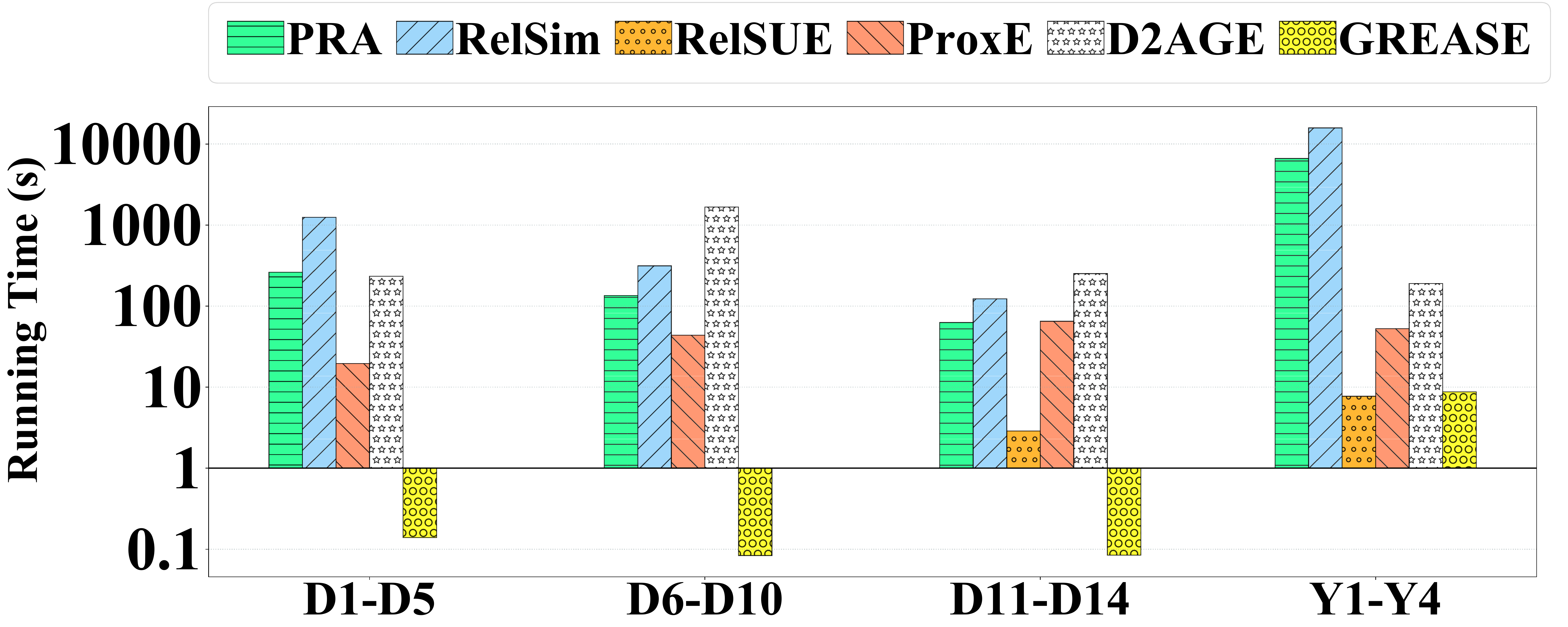}}
\caption{Average running time per query instance.}
\label{fig:timeall}
\end{figure}

\subsection{Supervised Relevance Search}

Relevance is rather ambiguous on a schema-rich KG. Unfortunately, users may not have the expertise to formally characterize the desired semantics of relevance, due to the complexity of the query language or the richness of the schema. In order to learn user-defined relevance, existing methods are mainly supervised by user-provided examples.
An early work is SRW~\cite{srw}, which leverages user-provided examples to supervise random walks for computing relevance.
In~\cite{ug,mpprefer}, predefined meta-paths are used to constrain random walks, and user-provided examples are used to learn the weights of the meta-paths.
However, it is unrealistic to predefine meta-paths for all possible types of information needs that users may have on a schema-rich KG. It is also difficult for a non-expert user to identify appropriate meta-paths from numerous candidates.

To tackle the problem, PRA~\cite{pra} and RelSim~\cite{relsim} automatically generate all possible meta-paths but they have to bound the length of an allowable meta-path because the number of possible meta-paths increases exponentially with length.
In~\cite{metagraph}, length-bounded meta-path is extended to size-bounded meta-graph.
In FSPG~\cite{fspg} and RelSUE~\cite{relsue}, there is no explicit length bound, but long meta-paths are penalized and hence are more likely to be pruned in their greedy search algorithms.
All these methods train a discriminative model to learn the weight of each meta-path or meta-graph from user-provided positive examples and randomly sampled negative examples. By contrast, we present a generative model which does not rely on negative examples. It outperforms the above discriminative methods in the experiments, in both effectiveness and efficiency. Moreover, our approach generalizes meta-paths into facets which also include properties that constrain answer entities, and hence it supports more expressive representation of user-defined relevance.

Some recent efforts learn graph embedding models for relevance~\cite{proxembed,d2age}.
However, they did not show better performance than other methods in our experiments. Their complex models may be more suitable for the link prediction task~\cite{lp1,lp2}, where large training sets are available. In relevance search, a user is not likely to provide many examples to supervise.

\begin{figure}[t]
\centerline{\includegraphics[width=\columnwidth]{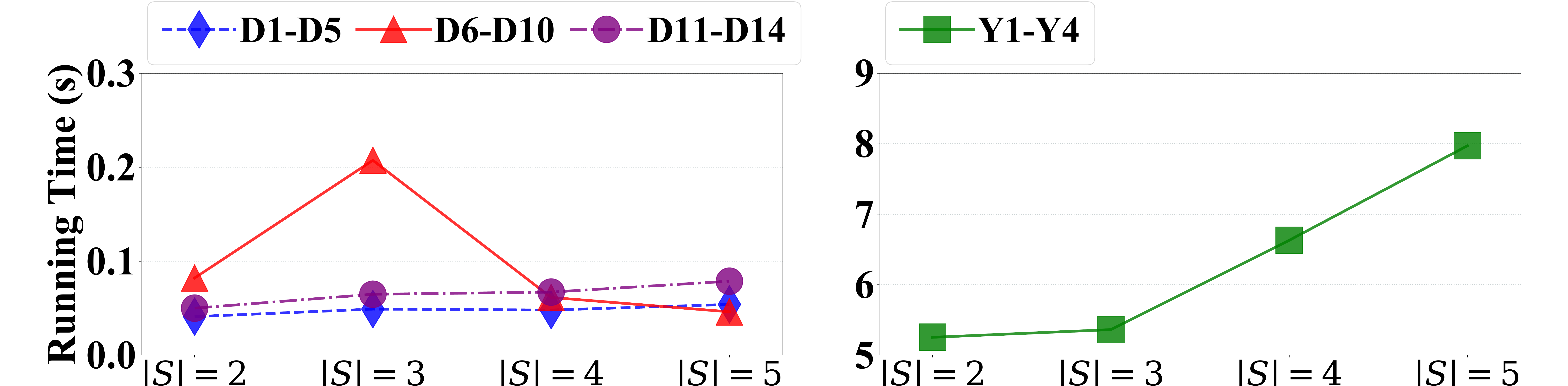}}
\caption{Average running time of GREASE per query instance.}
\label{fig:timeour}
\end{figure}

\subsection{Other Related Problems}

Other related problems include graph query by example~\cite{gqbe} and exemplar query answering~\cite{exemplar}, but their technical challenges and methods are fundamentally different. Their input is a tuple of entities~\cite{gqbe} or a keyword query~\cite{exemplar} provided by the user as an example. The example is expanded~\cite{gqbe} or mapped~\cite{exemplar} into a subgraph of the KG, called a query graph, to capture the user's query intent. The output is a set of other top-ranked subgraphs of the KG that are similar to the query graph. By contrast, relevance search is mainly focused on the selection, weighting, and combination of meta-paths to represent the user-defined relevance between the source entity and the target entity in the user-provided examples.

%% file: conclusion.tex
\section{Conclusions}\label{sect:con}
We proposed GREASE, a new approach to relevance search over KGs. Compared with existing methods, GREASE is distinguished by its more effective generative model for weighting meta-paths, its more expressive facet-based representation of relevance with properties, and its efficient implementation. These technical contributions have been demonstrated in an extensive evaluation.

One limitation of our approach is that our estimation of probabilities is largely based on frequency counts in the KG. However,
KGs in the real world
may be inexact or incomplete, which may affect the accuracy of our estimation. We have used smoothing methods to partially address this issue,
but further attempts may be helpful, e.g.,~using external knowledge.
In future work, we will also consider using ontological schemata and reasoning services to handle more complex semantics of relevance.


%% file: awsdm2020.bbl

\begin{thebibliography}{24}


\ifx \showCODEN    \undefined \def \showCODEN     #1{\unskip}     \fi
\ifx \showDOI      \undefined \def \showDOI       #1{#1}\fi
\ifx \showISBNx    \undefined \def \showISBNx     #1{\unskip}     \fi
\ifx \showISBNxiii \undefined \def \showISBNxiii  #1{\unskip}     \fi
\ifx \showISSN     \undefined \def \showISSN      #1{\unskip}     \fi
\ifx \showLCCN     \undefined \def \showLCCN      #1{\unskip}     \fi
\ifx \shownote     \undefined \def \shownote      #1{#1}          \fi
\ifx \showarticletitle \undefined \def \showarticletitle #1{#1}   \fi
\ifx \showURL      \undefined \def \showURL       {\relax}        \fi
\providecommand\bibfield[2]{#2}
\providecommand\bibinfo[2]{#2}
\providecommand\natexlab[1]{#1}
\providecommand\showeprint[2][]{arXiv:#2}

\bibitem[\protect\citeauthoryear{Backstrom and Leskovec}{Backstrom and
  Leskovec}{2011}]%
        {srw}
\bibfield{author}{\bibinfo{person}{Lars Backstrom} {and} \bibinfo{person}{Jure
  Leskovec}.} \bibinfo{year}{2011}\natexlab{}.
\newblock \showarticletitle{Supervised random walks: predicting and
  recommending links in social networks}. In \bibinfo{booktitle}{\emph{Proc.
  {WSDM}}}. \bibinfo{pages}{635--644}.
\newblock
\urldef\tempurl%
\url{https://doi.org/10.1145/1935826.1935914}
\showDOI{\tempurl}


\bibitem[\protect\citeauthoryear{Balmin, Hristidis, and
  Papakonstantinou}{Balmin et~al\mbox{.}}{2004}]%
        {objectrank}
\bibfield{author}{\bibinfo{person}{Andrey Balmin}, \bibinfo{person}{Vagelis
  Hristidis}, {and} \bibinfo{person}{Yannis Papakonstantinou}.}
  \bibinfo{year}{2004}\natexlab{}.
\newblock \showarticletitle{ObjectRank: Authority-Based Keyword Search in
  Databases}. In \bibinfo{booktitle}{\emph{Proc. {VLDB}}}.
  \bibinfo{pages}{564--575}.
\newblock


\bibitem[\protect\citeauthoryear{Bu, Hong, Peng, and Li}{Bu
  et~al\mbox{.}}{2014}]%
        {mpprefer}
\bibfield{author}{\bibinfo{person}{Shaoli Bu}, \bibinfo{person}{Xiaoguang
  Hong}, \bibinfo{person}{Zhaohui Peng}, {and} \bibinfo{person}{Qingzhong Li}.}
  \bibinfo{year}{2014}\natexlab{}.
\newblock \showarticletitle{Integrating meta-path selection with
  user-preference for top-k relevant search in heterogeneous information
  networks}. In \bibinfo{booktitle}{\emph{Proc. {CSCWD}}}.
  \bibinfo{pages}{301--306}.
\newblock
\urldef\tempurl%
\url{https://doi.org/10.1109/CSCWD.2014.6846859}
\showDOI{\tempurl}


\bibitem[\protect\citeauthoryear{Cai, Zheng, and Chang}{Cai
  et~al\mbox{.}}{2018}]%
        {lp1}
\bibfield{author}{\bibinfo{person}{HongYun Cai}, \bibinfo{person}{Vincent~W.
  Zheng}, {and} \bibinfo{person}{Kevin~Chen{-}Chuan Chang}.}
  \bibinfo{year}{2018}\natexlab{}.
\newblock \showarticletitle{A Comprehensive Survey of Graph Embedding:
  Problems, Techniques, and Applications}.
\newblock \bibinfo{journal}{\emph{{IEEE} Trans. Knowl. Data Eng.}}
  \bibinfo{volume}{30}, \bibinfo{number}{9} (\bibinfo{year}{2018}),
  \bibinfo{pages}{1616--1637}.
\newblock
\urldef\tempurl%
\url{https://doi.org/10.1109/TKDE.2018.2807452}
\showDOI{\tempurl}


\bibitem[\protect\citeauthoryear{Fang, Lin, Zheng, Wu, Chang, and Li}{Fang
  et~al\mbox{.}}{2016}]%
        {metagraph}
\bibfield{author}{\bibinfo{person}{Yuan Fang}, \bibinfo{person}{Wenqing Lin},
  \bibinfo{person}{Vincent~Wenchen Zheng}, \bibinfo{person}{Min Wu},
  \bibinfo{person}{Kevin~Chen{-}Chuan Chang}, {and} \bibinfo{person}{Xiaoli
  Li}.} \bibinfo{year}{2016}\natexlab{}.
\newblock \showarticletitle{Semantic proximity search on graphs with
  metagraph-based learning}. In \bibinfo{booktitle}{\emph{Proc. {ICDE}}}.
  \bibinfo{pages}{277--288}.
\newblock
\urldef\tempurl%
\url{https://doi.org/10.1109/ICDE.2016.7498247}
\showDOI{\tempurl}


\bibitem[\protect\citeauthoryear{Gu, Zhou, Cheng, Li, Pan, and Qu}{Gu
  et~al\mbox{.}}{2019}]%
        {relsue}
\bibfield{author}{\bibinfo{person}{Yu Gu}, \bibinfo{person}{Tianshuo Zhou},
  \bibinfo{person}{Gong Cheng}, \bibinfo{person}{Ziyang Li},
  \bibinfo{person}{Jeff~Z. Pan}, {and} \bibinfo{person}{Yuzhong Qu}.}
  \bibinfo{year}{2019}\natexlab{}.
\newblock \showarticletitle{Relevance Search over Schema-Rich Knowledge
  Graphs}. In \bibinfo{booktitle}{\emph{Proc. {WSDM}}}.
  \bibinfo{pages}{114--122}.
\newblock
\urldef\tempurl%
\url{https://doi.org/10.1145/3289600.3290970}
\showDOI{\tempurl}


\bibitem[\protect\citeauthoryear{Huang, Zheng, Cheng, Sun, Mamoulis, and
  Li}{Huang et~al\mbox{.}}{2016}]%
        {metastructure}
\bibfield{author}{\bibinfo{person}{Zhipeng Huang}, \bibinfo{person}{Yudian
  Zheng}, \bibinfo{person}{Reynold Cheng}, \bibinfo{person}{Yizhou Sun},
  \bibinfo{person}{Nikos Mamoulis}, {and} \bibinfo{person}{Xiang Li}.}
  \bibinfo{year}{2016}\natexlab{}.
\newblock \showarticletitle{Meta Structure: Computing Relevance in Large
  Heterogeneous Information Networks}. In \bibinfo{booktitle}{\emph{Proc.
  {SIGKDD}}}. \bibinfo{pages}{1595--1604}.
\newblock
\urldef\tempurl%
\url{https://doi.org/10.1145/2939672.2939815}
\showDOI{\tempurl}


\bibitem[\protect\citeauthoryear{Jayaram, Khan, Li, Yan, and Elmasri}{Jayaram
  et~al\mbox{.}}{2015}]%
        {gqbe}
\bibfield{author}{\bibinfo{person}{Nandish Jayaram}, \bibinfo{person}{Arijit
  Khan}, \bibinfo{person}{Chengkai Li}, \bibinfo{person}{Xifeng Yan}, {and}
  \bibinfo{person}{Ramez Elmasri}.} \bibinfo{year}{2015}\natexlab{}.
\newblock \showarticletitle{Querying Knowledge Graphs by Example Entity
  Tuples}.
\newblock \bibinfo{journal}{\emph{{IEEE} Trans. Knowl. Data Eng.}}
  \bibinfo{volume}{27}, \bibinfo{number}{10} (\bibinfo{year}{2015}),
  \bibinfo{pages}{2797--2811}.
\newblock
\urldef\tempurl%
\url{https://doi.org/10.1109/TKDE.2015.2426696}
\showDOI{\tempurl}


\bibitem[\protect\citeauthoryear{Lao and Cohen}{Lao and Cohen}{2010}]%
        {pra}
\bibfield{author}{\bibinfo{person}{Ni Lao} {and} \bibinfo{person}{William~W.
  Cohen}.} \bibinfo{year}{2010}\natexlab{}.
\newblock \showarticletitle{Relational retrieval using a combination of
  path-constrained random walks}.
\newblock \bibinfo{journal}{\emph{Machine Learning}} \bibinfo{volume}{81},
  \bibinfo{number}{1} (\bibinfo{year}{2010}), \bibinfo{pages}{53--67}.
\newblock
\urldef\tempurl%
\url{https://doi.org/10.1007/s10994-010-5205-8}
\showDOI{\tempurl}


\bibitem[\protect\citeauthoryear{Lehmann, Isele, Jakob, Jentzsch, Kontokostas,
  Mendes, Hellmann, Morsey, van Kleef, Auer, and Bizer}{Lehmann
  et~al\mbox{.}}{2015}]%
        {DBpedia}
\bibfield{author}{\bibinfo{person}{Jens Lehmann}, \bibinfo{person}{Robert
  Isele}, \bibinfo{person}{Max Jakob}, \bibinfo{person}{Anja Jentzsch},
  \bibinfo{person}{Dimitris Kontokostas}, \bibinfo{person}{Pablo~N. Mendes},
  \bibinfo{person}{Sebastian Hellmann}, \bibinfo{person}{Mohamed Morsey},
  \bibinfo{person}{Patrick van Kleef}, \bibinfo{person}{S{\"{o}}ren Auer},
  {and} \bibinfo{person}{Christian Bizer}.} \bibinfo{year}{2015}\natexlab{}.
\newblock \showarticletitle{DBpedia - {A} large-scale, multilingual knowledge
  base extracted from Wikipedia}.
\newblock \bibinfo{journal}{\emph{Semantic Web}} \bibinfo{volume}{6},
  \bibinfo{number}{2} (\bibinfo{year}{2015}), \bibinfo{pages}{167--195}.
\newblock
\urldef\tempurl%
\url{https://doi.org/10.3233/SW-140134}
\showDOI{\tempurl}


\bibitem[\protect\citeauthoryear{Liu, Zheng, Zhao, Zhu, Chang, Wu, and
  Ying}{Liu et~al\mbox{.}}{2017}]%
        {proxembed}
\bibfield{author}{\bibinfo{person}{Zemin Liu}, \bibinfo{person}{Vincent~W.
  Zheng}, \bibinfo{person}{Zhou Zhao}, \bibinfo{person}{Fanwei Zhu},
  \bibinfo{person}{Kevin~Chen{-}Chuan Chang}, \bibinfo{person}{Minghui Wu},
  {and} \bibinfo{person}{Jing Ying}.} \bibinfo{year}{2017}\natexlab{}.
\newblock \showarticletitle{Semantic Proximity Search on Heterogeneous Graph by
  Proximity Embedding}. In \bibinfo{booktitle}{\emph{Proc. {AAAI}}}.
  \bibinfo{pages}{154--160}.
\newblock


\bibitem[\protect\citeauthoryear{Liu, Zheng, Zhao, Zhu, Chang, Wu, and
  Ying}{Liu et~al\mbox{.}}{2018}]%
        {d2age}
\bibfield{author}{\bibinfo{person}{Zemin Liu}, \bibinfo{person}{Vincent~W.
  Zheng}, \bibinfo{person}{Zhou Zhao}, \bibinfo{person}{Fanwei Zhu},
  \bibinfo{person}{Kevin~Chen{-}Chuan Chang}, \bibinfo{person}{Minghui Wu},
  {and} \bibinfo{person}{Jing Ying}.} \bibinfo{year}{2018}\natexlab{}.
\newblock \showarticletitle{Distance-Aware {DAG} Embedding for Proximity Search
  on Heterogeneous Graphs}. In \bibinfo{booktitle}{\emph{Proc. {AAAI}}}.
\newblock


\bibitem[\protect\citeauthoryear{Mahdisoltani, Biega, and
  Suchanek}{Mahdisoltani et~al\mbox{.}}{2015}]%
        {YAGO}
\bibfield{author}{\bibinfo{person}{Farzaneh Mahdisoltani},
  \bibinfo{person}{Joanna Biega}, {and} \bibinfo{person}{Fabian~M. Suchanek}.}
  \bibinfo{year}{2015}\natexlab{}.
\newblock \showarticletitle{{YAGO3:} {A} Knowledge Base from Multilingual
  Wikipedias}. In \bibinfo{booktitle}{\emph{Proc. {CIDR}}}.
\newblock


\bibitem[\protect\citeauthoryear{Mart{\'{\i}}nez, Berzal, and
  Talavera}{Mart{\'{\i}}nez et~al\mbox{.}}{2017}]%
        {lp2}
\bibfield{author}{\bibinfo{person}{V{\'{\i}}ctor Mart{\'{\i}}nez},
  \bibinfo{person}{Fernando Berzal}, {and} \bibinfo{person}{Juan Carlos~Cubero
  Talavera}.} \bibinfo{year}{2017}\natexlab{}.
\newblock \showarticletitle{A Survey of Link Prediction in Complex Networks}.
\newblock \bibinfo{journal}{\emph{{ACM} Comput. Surv.}} \bibinfo{volume}{49},
  \bibinfo{number}{4} (\bibinfo{year}{2017}), \bibinfo{pages}{69:1--69:33}.
\newblock
\urldef\tempurl%
\url{https://doi.org/10.1145/3012704}
\showDOI{\tempurl}


\bibitem[\protect\citeauthoryear{Meng, Cheng, Maniu, Senellart, and Zhang}{Meng
  et~al\mbox{.}}{2015}]%
        {fspg}
\bibfield{author}{\bibinfo{person}{Changping Meng}, \bibinfo{person}{Reynold
  Cheng}, \bibinfo{person}{Silviu Maniu}, \bibinfo{person}{Pierre Senellart},
  {and} \bibinfo{person}{Wangda Zhang}.} \bibinfo{year}{2015}\natexlab{}.
\newblock \showarticletitle{Discovering Meta-Paths in Large Heterogeneous
  Information Networks}. In \bibinfo{booktitle}{\emph{Proc. {WWW}}}.
  \bibinfo{pages}{754--764}.
\newblock
\urldef\tempurl%
\url{https://doi.org/10.1145/2736277.2741123}
\showDOI{\tempurl}


\bibitem[\protect\citeauthoryear{Mottin, Lissandrini, Velegrakis, and
  Palpanas}{Mottin et~al\mbox{.}}{2016}]%
        {exemplar}
\bibfield{author}{\bibinfo{person}{Davide Mottin}, \bibinfo{person}{Matteo
  Lissandrini}, \bibinfo{person}{Yannis Velegrakis}, {and}
  \bibinfo{person}{Themis Palpanas}.} \bibinfo{year}{2016}\natexlab{}.
\newblock \showarticletitle{Exemplar queries: a new way of searching}.
\newblock \bibinfo{journal}{\emph{{VLDB} J.}} \bibinfo{volume}{25},
  \bibinfo{number}{6} (\bibinfo{year}{2016}), \bibinfo{pages}{741--765}.
\newblock
\urldef\tempurl%
\url{https://doi.org/10.1007/s00778-016-0429-2}
\showDOI{\tempurl}


\bibitem[\protect\citeauthoryear{Shi, Kong, Yu, Xie, and Wu}{Shi
  et~al\mbox{.}}{2012}]%
        {hetesim}
\bibfield{author}{\bibinfo{person}{Chuan Shi}, \bibinfo{person}{Xiangnan Kong},
  \bibinfo{person}{Philip~S. Yu}, \bibinfo{person}{Sihong Xie}, {and}
  \bibinfo{person}{Bin Wu}.} \bibinfo{year}{2012}\natexlab{}.
\newblock \showarticletitle{Relevance search in heterogeneous networks}. In
  \bibinfo{booktitle}{\emph{Proc. {EDBT}}}. \bibinfo{pages}{180--191}.
\newblock
\urldef\tempurl%
\url{https://doi.org/10.1145/2247596.2247618}
\showDOI{\tempurl}


\bibitem[\protect\citeauthoryear{Shi, Chan, Zhuang, Gui, and Han}{Shi
  et~al\mbox{.}}{2017}]%
        {prep}
\bibfield{author}{\bibinfo{person}{Yu Shi}, \bibinfo{person}{Po{-}Wei Chan},
  \bibinfo{person}{Honglei Zhuang}, \bibinfo{person}{Huan Gui}, {and}
  \bibinfo{person}{Jiawei Han}.} \bibinfo{year}{2017}\natexlab{}.
\newblock \showarticletitle{{PReP}: Path-Based Relevance from a Probabilistic
  Perspective in Heterogeneous Information Networks}. In
  \bibinfo{booktitle}{\emph{Proc. {SIGKDD}}}. \bibinfo{pages}{425--434}.
\newblock
\urldef\tempurl%
\url{https://doi.org/10.1145/3097983.3097990}
\showDOI{\tempurl}


\bibitem[\protect\citeauthoryear{Sun, Barber, Gupta, Aggarwal, and Han}{Sun
  et~al\mbox{.}}{2011a}]%
        {pc}
\bibfield{author}{\bibinfo{person}{Yizhou Sun}, \bibinfo{person}{Rick Barber},
  \bibinfo{person}{Manish Gupta}, \bibinfo{person}{Charu~C. Aggarwal}, {and}
  \bibinfo{person}{Jiawei Han}.} \bibinfo{year}{2011}\natexlab{a}.
\newblock \showarticletitle{Co-author Relationship Prediction in Heterogeneous
  Bibliographic Networks}. In \bibinfo{booktitle}{\emph{Proc. {ASONAM}}}.
  \bibinfo{pages}{121--128}.
\newblock
\urldef\tempurl%
\url{https://doi.org/10.1109/ASONAM.2011.112}
\showDOI{\tempurl}


\bibitem[\protect\citeauthoryear{Sun, Han, Yan, Yu, and Wu}{Sun
  et~al\mbox{.}}{2011b}]%
        {pathsim}
\bibfield{author}{\bibinfo{person}{Yizhou Sun}, \bibinfo{person}{Jiawei Han},
  \bibinfo{person}{Xifeng Yan}, \bibinfo{person}{Philip~S. Yu}, {and}
  \bibinfo{person}{Tianyi Wu}.} \bibinfo{year}{2011}\natexlab{b}.
\newblock \showarticletitle{{PathSim}: Meta Path-Based Top-K Similarity Search
  in Heterogeneous Information Networks}.
\newblock \bibinfo{journal}{\emph{{PVLDB}}} \bibinfo{volume}{4},
  \bibinfo{number}{11} (\bibinfo{year}{2011}), \bibinfo{pages}{992--1003}.
\newblock


\bibitem[\protect\citeauthoryear{Sun, Norick, Han, Yan, Yu, and Yu}{Sun
  et~al\mbox{.}}{2012}]%
        {PathSelClus}
\bibfield{author}{\bibinfo{person}{Yizhou Sun}, \bibinfo{person}{Brandon
  Norick}, \bibinfo{person}{Jiawei Han}, \bibinfo{person}{Xifeng Yan},
  \bibinfo{person}{Philip~S. Yu}, {and} \bibinfo{person}{Xiao Yu}.}
  \bibinfo{year}{2012}\natexlab{}.
\newblock \showarticletitle{Integrating meta-path selection with user-guided
  object clustering in heterogeneous information networks}. In
  \bibinfo{booktitle}{\emph{Proc. {SIGKDD}}}. \bibinfo{pages}{1348--1356}.
\newblock
\urldef\tempurl%
\url{https://doi.org/10.1145/2339530.2339738}
\showDOI{\tempurl}


\bibitem[\protect\citeauthoryear{Wang, Sun, Song, Han, Song, Wang, and
  Zhang}{Wang et~al\mbox{.}}{2016}]%
        {relsim}
\bibfield{author}{\bibinfo{person}{Chenguang Wang}, \bibinfo{person}{Yizhou
  Sun}, \bibinfo{person}{Yanglei Song}, \bibinfo{person}{Jiawei Han},
  \bibinfo{person}{Yangqiu Song}, \bibinfo{person}{Lidan Wang}, {and}
  \bibinfo{person}{Ming Zhang}.} \bibinfo{year}{2016}\natexlab{}.
\newblock \showarticletitle{RelSim: Relation Similarity Search in Schema-Rich
  Heterogeneous Information Networks}. In \bibinfo{booktitle}{\emph{Proc.
  {SDM}}}. \bibinfo{pages}{621--629}.
\newblock
\urldef\tempurl%
\url{https://doi.org/10.1137/1.9781611974348.70}
\showDOI{\tempurl}


\bibitem[\protect\citeauthoryear{Xiong, Zhu, and Yu}{Xiong
  et~al\mbox{.}}{2015}]%
        {joinsim}
\bibfield{author}{\bibinfo{person}{Yun Xiong}, \bibinfo{person}{Yangyong Zhu},
  {and} \bibinfo{person}{Philip~S. Yu}.} \bibinfo{year}{2015}\natexlab{}.
\newblock \showarticletitle{Top-k Similarity Join in Heterogeneous Information
  Networks}.
\newblock \bibinfo{journal}{\emph{{IEEE} Trans. Knowl. Data Eng.}}
  \bibinfo{volume}{27}, \bibinfo{number}{6} (\bibinfo{year}{2015}),
  \bibinfo{pages}{1710--1723}.
\newblock
\urldef\tempurl%
\url{https://doi.org/10.1109/TKDE.2014.2373385}
\showDOI{\tempurl}


\bibitem[\protect\citeauthoryear{Yu, Sun, Norick, Mao, and Han}{Yu
  et~al\mbox{.}}{2012}]%
        {ug}
\bibfield{author}{\bibinfo{person}{Xiao Yu}, \bibinfo{person}{Yizhou Sun},
  \bibinfo{person}{Brandon Norick}, \bibinfo{person}{Tiancheng Mao}, {and}
  \bibinfo{person}{Jiawei Han}.} \bibinfo{year}{2012}\natexlab{}.
\newblock \showarticletitle{User guided entity similarity search using
  meta-path selection in heterogeneous information networks}. In
  \bibinfo{booktitle}{\emph{Proc. {CIKM}}}. \bibinfo{pages}{2025--2029}.
\newblock
\urldef\tempurl%
\url{https://doi.org/10.1145/2396761.2398565}
\showDOI{\tempurl}


\end{thebibliography}
